\documentstyle[12pt]{article}

\newcommand{\be}{\begin{equation}}
\newcommand{\ee}{\end{equation}}
\begin{document}

\thispagestyle{empty}

\begin{flushright}
\vspace{1mm}
FIAN/TD/20--96\\
{August 1996}\\
\end{flushright}
\vspace{1cm}

\begin{center}
{\large\bf
FREE EQUATIONS FOR MASSIVE MATTER FIELDS IN 2+1 -DIMENSIONAL
ANTI-DE SITTER SPACE
{}From DEFORMED OSCILLATOR ALGEBRA}

\vglue 1.5 true cm

{\bf A.~V.~BARABANSCHIKOV,}\\

\vglue 0.3 true cm

International School for Advanced Studies \\
Via Beirut 2-4, 34013, Trieste, Italy \\

\vglue 1 true cm

{\bf S.~F.~PROKUSHKIN, AND
M.~A.~VASILIEV,}

\vglue 0.3 true cm

I.E.Tamm Department of Theoretical Physics, Lebedev Physical Institute,\\
Leninsky prospect 53, 117924, Moscow, Russia
\medskip
\end{center}

\vglue 3 true cm

\begin{abstract}
\baselineskip .4 true cm
\noindent
We reformulate free equations of motion for massive spin 0 and spin 1/2
matter fields in 2+1 dimensional anti-de Sitter space
in the form of some covariant constantness conditions.
The infinite-dimensional
representation of the anti-de Sitter algebra
underlying this formulation is shown to admit a natural
realization in terms of the algebra of deformed oscillators with a
deformation parameter related to the parameter of mass.
\end{abstract}

\newpage

\section{Introduction}

In \cite{Unf} free equations of spin 0 and spin 1/2 matter fields
in 2+1 - dimensional anti-de Sitter (AdS) space were reformulated
in a form of certain covariant constantness
conditions (``unfolded form''). Being equivalent to the standard
one, such a formulation is useful at least in two respects. It
leads to a simple construction of a general solution of the free
equations and gives important hints how to describe
non-linear dynamics exhibiting infinite-dimensional higher-spin
symmetries. In \cite{Unf} it was also observed that the
proposed construction admits a natural realization in terms of
the Heisenberg-Weyl oscillator algebra for the case of massless
fields. Based on this realization, non-linear dynamics of
massless matter fields interacting through higher-spin gauge
fields was then formulated in \cite{Eq} in all orders in
interactions.

In the present paper we address the question how one can
extend the oscillator realization of
the massless equations of \cite{Unf} to the case of
an arbitrary mass of matter fields. We show that
the relevant algebraic construction is provided
by the deformed oscillator algebra suggested in \cite{Quant}
with the deformation parameter related to the parameter of mass.
In a future publication of the two of the authors \cite{Fut}
the results of this paper will be used
for the analysis of non-linear dynamics of matter fields
in 2+1 dimensions, interacting through higher-spin gauge fields.
The 2+1 dimensional model considered in this
paper can be regarded as a toy model exhibiting some of the general properties
of physically more important higher-spin gauge
theories
in higher dimensions $d\geq 4$.

\section{Preliminaries}
 We describe the 2+1 dimensional
AdS space in terms of
the Lorentz connection one-form
$\omega^{\alpha\beta}=dx^\nu
\omega_\nu{}^{\alpha\beta}(x)$
and dreibein one-form
$h^{\alpha\beta}=
dx^\nu h_\nu{}^{\alpha\beta} (x)$.
Here
$x^\nu$ are
space-time coordinates
$(\nu =0,1,2)$
and
$\alpha,\beta,\ldots =1,2$ are spinor indices, which are
raised and lowered with the aid of the symplectic form
$\epsilon_{\alpha\beta}=-\epsilon_{\beta\alpha}$,
$A^{\alpha}=\epsilon^{\alpha\beta}A_{\beta}$,
$A_{\alpha}=A^{\beta}\epsilon_{\beta\alpha}$,
$\epsilon_{12}=\epsilon^{12}=1$.
The AdS geometry can be
described by the equations
\be
\label{d omega}
    d\omega_{\alpha\beta}=\omega_{\alpha\gamma}\wedge\omega_\beta{}^\gamma+
\lambda^2h_{\alpha\gamma}\wedge h_\beta{}^\gamma\,,
\ee
\be
\label{dh}
    dh_{\alpha\beta}=\omega_{\alpha\gamma}\wedge h_\beta{}^\gamma+
  \omega_{\beta\gamma}\wedge h_\alpha{}^\gamma \, ,
\ee
which have a form of zero-curvature conditions for the
$o(2,2)\sim sp(2)\oplus sp(2)$ Yang-Mills field strengths.
Here $\omega_{\alpha\beta}$ and $h_{\alpha\beta}$ are
symmetric in $\alpha$ and $\beta$.
For the space-time geometric interpretation of these equations
one has to assume that the dreibein $h_\nu{}^{\alpha\beta}$ is
a non-degenerate $3\times 3$ matrix.
Then
(\ref{dh}) reduces to the zero-torsion condition which expresses
Lorentz connection via dreibein $h_\nu{}^{\alpha\beta}$ and (\ref{d omega})
implies that the Riemann tensor 2-form $R_{\alpha\beta}=
    d\omega_{\alpha\beta}-\omega_{\alpha\gamma}\wedge\omega_\beta{}^\gamma$
acquires the AdS form
\be
\label{R}
R_{\alpha\beta}=
\lambda^2h_{\alpha\gamma}\wedge h_\beta{}^\gamma
\ee
with
$\lambda^{-1}$  identified with the AdS radius.

In \cite{Unf} it was shown that
one can reformulate free field equations for matter fields in
2+1 dimensions
in terms of the generating function
$C(y|x)$
\be
\label{C0}
C(y|x)=
\sum_{n=0}^\infty \frac1{n!}C_{\alpha_1 \ldots\alpha_n}(x) y^{\alpha_1}\ldots
      y^{\alpha_n}\,
\ee
in the following ``unfolded'' form
\be
\label{DC mod}
  DC=h^{\alpha\beta} \left[a(N)
  \frac{\partial}{\partial y^\alpha }\frac{\partial}{\partial y^\beta}+
  b(N) y_\alpha\frac{\partial}{\partial y^\beta}+ e(N)
  y_\alpha y_\beta \right]C \, ,
\ee
where $D$ is the Lorentz covariant differential
\be
\label{lorcov}
   D=d-\omega^{\alpha\beta}y_\alpha \frac{\partial}{\partial y^\beta}\,
\ee
and $N$ is the Euler operator
\be
N\equiv y^\alpha\frac{\partial}{\partial y^\alpha} \, .
\ee
The integrability conditions of the equations (\ref{DC mod})
(i.e. the consistency with $d^2 =0$) require
the functions $a,b$ and $e$ to satisfy the following restrictions \cite{Unf}
\be
\label{consist}
  \alpha(n)=0 \mbox{\qquad for \, $n\ge 0$ ,\qquad $\gamma(n)=0$
  \qquad for \quad $n\ge 2$ ,}
\ee
$$
  \beta(n)=0\mbox{ \qquad for \quad  $n\ge 1$ ,}
$$
where
\be
  \alpha(N)=a(N)\left[(N+4)b(N+2)-Nb(N)\right]\,,
\ee
\be
  \gamma(N)=e(N)\left[(N+2)b(N)-(N-2)b(N-2)\right]\,,
\ee
\be
  \beta(N)=(N+3)a(N)e(N+2)-(N-1)e(N)a(N-2)+b^2(N)-\lambda^2\, .
\ee
It was shown in \cite{Unf} that, for the condition that
$a(n)\ne 0$ $\forall$ $n\ge 0$ and up to a freedom of
field redefinitions $C\rightarrow \tilde{C} =\varphi (N) C$,
$\varphi(n) \neq 0 \quad \forall n\in {\bf Z}^{+}$,
there exist two one parametric
classes of independent solutions of~(\ref{consist}),
$$
   a(n)=1\,,\qquad b(n)=0\,, \qquad e(n)=\frac14\lambda^2-\frac{M^2}
   {2(n+1)(n-1)}\, ,\qquad
n\,\mbox{--even}\,,
$$
\be
\label{cob}
    a(n)=b(n)=e(n)=0\,,\qquad n\,\mbox{--odd}\,,
\ee
and
$$
    a(n)=b(n)=e(n)=0\,,\qquad n\,\mbox{--even}\,,
$$
\be
\label{cof}
   a(n)=1\,,\qquad b(n)=\frac{\sqrt2M}{n(n+2)}\,,\qquad
   e(n)=\frac14\lambda^2-\frac{M^2}{2n^2}\, ,\qquad
     n\,\mbox{--odd}\,,
\ee
with an arbitrary parameter $M$.
As a result, the system (\ref{DC mod})
reduces to two independent infinite chains of equations
for bosons and fermions described by multispinors with even and odd number
of indices, respectively.
To elucidate the physical content of these equations
 one has to identify the
lowest components of the expansion (\ref{C0}), $C(x)$ and $C_\alpha (x)$,
with the physical spin-0 boson and spin 1/2 fermion  matter fields,
 respectively, and to check,
first, that the system (\ref{DC mod}) amounts to the physical massive
Klein-Gordon and Dirac equations,
\be
\label{M K-G}
  \Box C=\left(\frac32\lambda^2-M^2\right)C\,,
\ee
\be
\label{D}
  h^\nu{}_\alpha{}^\beta D_{\nu}C_\beta=\frac M{\sqrt2} C_\alpha \,,
\ee
and, second, that
all other equations in (\ref{DC mod}) express
all highest multispinors via highest derivatives of
the matter fields $C$ and $C_{\alpha}$ imposing no additional
constraints on the latter. Note that the
D'Alambertian is defined as usual
\be
\label{dal}
 \Box =D^{\mu}D_{\mu} \,,
\ee       
where $D_{\mu}$ is
a full background covariant derivative
involving the zero-torsion Christoffel connection defined
through the metric postulate $D_{\mu}h_{\nu}^{\alpha\beta}=0$.
The inverse dreibein $h^\nu{}_{\alpha\beta}$ is defined as
in \cite{Unf},
\be
   h_\nu{}^{\alpha\beta}h^\nu{}_{\gamma\delta}=
    \frac12(\delta^\alpha_\gamma\delta^\beta_\delta+\delta^\alpha_
    \delta\delta^\beta_\gamma)\,.
\ee
Note also that the indices $\mu$, $\nu$ are raised and lowered
by the metric tensor
$$
   g_{\mu\nu}=h_\mu{}^{\alpha\beta}h_\nu{}_{\alpha\beta} \,.
$$

As emphasized in \cite{Unf}, the equations (\ref{DC mod})
provide a particular example of covariant constantness conditions
\be
\label{dC}
  dC_i =A_i{}^j C_j
\ee
with the gauge fields $A_i{}^j =A^a(T_a)_i{}^j$
obeying the zero-curvature conditions
\be
\label{dA}
  dA^a=U^a_{bc}A^b \wedge A^c \,,
\ee
where $U^a_{bc}$ are structure coefficients of the Lie (super)algebra
which gives rise to the gauge fields $A^a$ (cf (1), (2)).
Then the requirement that the integrability conditions
for (\ref{dC}) must be true is equivalent to the requirement that
$(T_a)_i{}^j$ form some matrix representation of the gauge algebra.
Thus, the problem consists of finding an appropriate representation of
the space-time symmetry group which leads to correct field equations.
As a result, after the equations are rewritten in this ``unfolded form'',
one can write down their general solution in a pure gauge form
$A(x)=-g^{-1}(x) dg(x)$, $C(x)=T(g^{-1})(x) C_0$, where  $C_0$ is
an arbitrary $x$ - independent element of the representation space.
This general solution has a structure of the covariantized Taylor
type expansion \cite{Unf}. For the problem under consideration the relevant
(infinite-dimensional) representation of the AdS algebra is characterized
by the coefficients (\ref{cob}) and (\ref{cof}).

\section{Operator Realization for Arbitrary Mass}

Let us now describe an operator algebra that leads automatically
to the correct massive field equations of the form
(\ref{DC mod}).

Following to \cite{Quant} we introduce oscillators obeying the commutation
relations
\be
\label{y mod}
  [\hat{y}_\alpha,\hat{y}_\beta]=2i\epsilon_{\alpha\beta}(1+\nu k)\, ,
\ee
where $\alpha ,\beta =1,2$, $k$ is the Klein operator anticommuting with
$\hat{y}_\alpha$,
\be
\label{k}
  k\hat{y}_\alpha=-\hat{y}_\alpha k\, ,  \qquad k^2 =1
\ee
and $\nu$ is a free parameter.
The main property of these oscillators is that
the bilinears
\be
\label{Q}
   T_{\alpha\beta} =\frac{1}{4i} \{\hat{y}_\alpha ,\hat{y}_\beta\}
\ee
fulfill the standard $sp(2)$ commutation relations
\be
\label{sp(2) com}
  [T_{\alpha\beta},T_{\gamma\delta}]=
    \epsilon_{\alpha\gamma}T_{\beta\delta}+
    \epsilon_{\beta\delta}T_{\alpha\gamma}+
    \epsilon_{\alpha\delta}T_{\beta\gamma}+
    \epsilon_{\beta\gamma}T_{\alpha\delta}
\ee
as well as
\be
\label{oscom}
  [T_{\alpha\beta} ,\hat{y}_{\gamma}]=
    \epsilon_{\alpha\gamma}\hat{y}_{\beta}+
    \epsilon_{\beta\gamma}\hat{y}_{\alpha}\,
\ee
for any $\nu$.
Note that a specific realization of this kind of oscillators
was considered by Wigner
\cite{Wig} who addressed a question whether it is possible to
modify the oscillator commutation relations in such a way that
the relation $[H, a_\pm ]=\pm a_\pm $ remains valid.
This relation is a particular case of (\ref{oscom}) with
$H=T_{12}$ and $a_\pm =y_{1,2}$.

The property
(\ref{sp(2) com})
 allows us to realize the $o(2,2)$ gravitational
 fields as
\be
\label{W}
  W_{gr} (x)= \omega +\lambda h ;\qquad
    \omega\equiv\frac1{8i}\omega^{\alpha\beta}\{\hat{y}_\alpha,
    \hat{y}_\beta\} \, ,
    \quad h\equiv\frac1{8i}h^{\alpha\beta}\{\hat{y}_\alpha,\hat{y}_\beta\}
    \psi \, ,
\ee
where $\psi$ is an additional central involutive element,
\be
   \psi^2=1\,,\qquad [\psi,\hat{y}_{\alpha}]=0\,,\qquad
   [\psi,k]=0\,,
\ee
which is introduced to describe the 3d AdS algebra
$o(2,2)\sim sp(2)\oplus sp(2)$ spanned by the generators
\be
\label{al}
   L_{\alpha\beta}=\frac1{4i}\{\hat{y}_\alpha,
    \hat{y}_\beta\}\,,\qquad \,
   P_{\alpha\beta}=\frac1{4i}\{\hat{y}_\alpha,
    \hat{y}_\beta\}\psi\,.
\ee
Now the equations (\ref{d omega}) and (\ref{dh})
describing the vacuum anti-de Sitter geometry acquire a
form
\be
\label{va}
   dW_{gr} =W_{gr} \wedge W_{gr}\, .
\ee

Let us introduce the operator-valued generating function $C(\hat{y},k|x)$
\be
\label{hatC}
  C(\hat{y},k,\psi|x)=\sum_{A,B=0,1}
  \sum_{n=0}^\infty \frac 1{n!} \lambda^{-[\frac n2]}
  C^{AB}_{\alpha_1 \ldots\alpha_n}(x) k^A
  \psi^B\hat{y}^{\alpha_1}\ldots \hat{y}^{\alpha_n}\, ,
\ee
where $C^{AB}_{\alpha_1 \ldots\alpha_n}$ are totally symmetric tensors
(which implies the Weyl ordering with respect to $\hat{y}_{\alpha}$).
It is easy to see that  the following two types of equations
\be
\label{aux}
    DC=\lambda[h,C] \, ,
\ee
and
\be
\label{D hatC}
    DC=\lambda\{h,C\} \, ,
\ee
where
\be
   DC\equiv dC-[\omega,C] \,
\ee
are consistent (i.e. the integrability conditions are satisfied
as a consequence of the vacuum conditions (\ref{va})). Indeed,
(\ref{aux}) corresponds to the adjoint action of the
space-time algebra (\ref{al}) on the algebra of modified
oscillators.  The equations (\ref{D hatC}) correspond to another
representation of the space-time symmetry which we call twisted
representation. The fact that one can replace the commutator
by the anticommutator in the term proportional to dreibein is a simple
consequence of the property that AdS algebra possesses an involutive
automorphism changing a sign of the AdS translations.
In the particular realization used here it is induced by the
automorphism $\psi\to -\psi$.

There is an important difference  between these two representations.
The first one involving the commutator decomposes into an infinite
direct sum of finite-dimensional representations of the space-time symmetry
algebra. Moreover, because of the property
(\ref{oscom}) this representation is $\nu$-independent and therefore is
equivalent to the representation with $\nu=0$ which was shown
in \cite{Unf} to describe
an infinite set of auxiliary (topological) fields. The twisted representation
on the other hand is just the infinite-dimensional representation
needed for the description of
matter fields (in what follows we will use the symbol $C$ only for the twisted
representation).

To see this one has to carry out a component analysis
of the equations
(\ref{D hatC}) which consists of some operator reorderings bringing
all terms into the Weyl ordered form with respect to
$\hat{y}_\alpha$.
As a result one finds that
(\ref{D hatC})
takes the form of the
equation (\ref{DC mod}) with the following
values of the coefficients $a(n)$, $b(n)$ and $e(n)$ :
\begin{eqnarray}
\label{a}
  \lefteqn{a(n)=\frac{i\lambda}2 \left[1+\nu k\frac{1+(-1)^n}{(n+2)^2-1}
    \right.} \nonumber\\
  & & \left.{}-\frac{\nu^2}{(n+2)^2((n+2)^2-1)} \left((n+2)^2-
  \frac{1-(-1)^n}2 \right)\right]\,,
\end{eqnarray}
\be
\label{b}
  b(n)=-\nu k\lambda\,\frac{1-(-1)^n}{2n(n+2)}\,,
\ee
\be
\label{e}
  e(n)=-\frac{i\lambda}2\, .
\ee
As expected, these expressions satisfy the conditions~(\ref{consist}).

Now let us remind ourselves that due to the presence of the Klein operator
$k$ we have a doubled number of fields compared to the analysis in the
beginning of this section. One can project out the irreducible
subsets with the aid of the two projectors $P_\pm$,
\be
  C_\pm\equiv P_\pm C\, ,\qquad P_\pm\equiv\frac{1\pm k}2\, .
\ee
As a result we get the following component form of eq.~(\ref{DC mod})
with the coefficients (\ref{a})-(\ref{e}),
\be
\label{chainbos+-}
   DC^{\pm}_{\alpha(n)}=\frac i2\left[\left(1-\frac{\nu(\nu\mp2)}{(n+1)(n+3)}
   \right) h^{\beta\gamma}C^{\pm}_{\beta\gamma\alpha(n)}-
   \lambda^2n(n-1)h_{\alpha\alpha}C^{\pm}_{\alpha(n-2)}\right]
\ee
for even $n$, and
\begin{eqnarray}
\label{chainferm+-}
  DC^{\pm}_{\alpha(n)} & = & \frac i2\left(1-\frac{\nu^2}{(n+2)^2}\right)
     h^{\beta\gamma}C^{\pm}_{\beta\gamma\alpha(n)} \pm
     \frac {\nu\lambda}{n+2}h_{\alpha}{}^{\beta}C^{\pm}_{\beta\alpha(n-1)}
     \nonumber\\
   & & {}-\frac i2 \lambda^2n(n-1)h_{\alpha\alpha}C^{\pm}_{\alpha(n-2)}
\end{eqnarray}
for odd $n$. Here we use the notation
$C_{\alpha(n)}=C_{\alpha_1,\dots,\alpha_n}$ and assume the full
symmetrization of the indices denoted by $\alpha$.

As it was shown in \cite{Unf}, the D'Alambertian
corresponding to eq.~(\ref{DC mod}) has the following form
\begin{eqnarray}
\label{D'Al}
  \Box C & = & \Biggl[(N+3)(N+2)a(N)e(N+2)+ \nonumber\\
  & & \left.+N(N-1)e(N)a(N-2)-\frac12N(N+2)b^2(N)\right]C\, .
\end{eqnarray}

Insertion of~(\ref{a})-(\ref{e}) into~(\ref{D'Al}) yields
\be
\label{L M}
  \Box C_\pm =\left[\lambda^2\frac{N(N+2)}2+\lambda^2\frac32-
  M^2_\pm \right]C_\pm\,,     
\ee
with
\be
\label{M}
  M^2_\pm =\lambda^2\frac{\nu(\nu\mp 2)}2\, ,\qquad n\mbox{ -even,}
\ee
\be
\label{M f}
  M^2_\pm =\lambda^2\frac{\nu^2}2\, ,\qquad n\mbox{ -odd.}
\ee

Thus, it is shown that the
modification (\ref{y mod}) allows one to
describe matter fields
\footnote{Let us remind the reader that the physical matter
field components are singled out by the conditions $NC_\pm=0$
in the bosonic sector and $NC_\pm=C_\pm$ in the fermionic sector}
with an arbitrary mass parameter related to $\nu$.
This construction generalizes in a natural way the realization
of equations for massless matter fields in terms of
the ordinary ($\nu=0$) oscillators proposed
in \cite{Unf}. An important comment however is that
this construction
not necessarily leads to non-vanishing coefficients $a(n)$.
Consider, for example,
expression~(\ref{a}) for the bosonic part of
$C_{+}$, i.e., set $k=1\,,\,n=2m$, $m$ is some integer,
\be
\label{a1}
   a(2m)=\frac{i\lambda}2 \left[1-\frac{\nu(\nu-2)}{(2m+1)(2m+3)}\right]\, .
\ee
We observe that $a(2l)=0$ at  $\nu=\pm 2(l+1)+1 $. It is not difficult to see
that some of the coefficients $a(n)$ vanish if and only if
$\nu=2k+1$ for some integer $k$.
This conclusion is in agreement with the results
of~\cite{Quant} where it was shown that for these values
of $\nu$ the enveloping algebra of the relations (\ref{y mod}),
$Aq(2;\nu |{\bf C})$, possesses ideals.
Thus, strictly speaking for $\nu =2k+1$
the system of equation derived from the operator realization
(\ref{D hatC}) is different from that considered in \cite{Unf}.
The specificities of the degenerated systems with
$\nu=2k+1$ will be discussed in the section 5.

In \cite{BWV} it was shown that the algebra $Aq(2,\nu )$
is isomorphic to the factor algebra $U(osp(1,2))/I(C_2 -\nu^2 )$,
where $U(osp(1,2))$ is the enveloping algebra
of $osp(1,2)$, while $I(C_2 -\nu^2 )$ is the ideal
spanned by all elements of the form
$$
  (C_2-\nu^2)\, x\,, \qquad \forall x\in U(osp(1,2)) \,,
$$
where $C_2$ is the quadratic Casimir operator of $osp(1,2)$.
{}From this observation it follows in particular that
the oscillator realization described above
is explicitly supersymmetric. In fact it is N=2 supersymmetric \cite{BWV}
with the generators of $osp(2,2)$ of the form
$$
  T_{\alpha\beta}=\frac1{4i}\{\hat{y}_\alpha,\hat{y}_\beta \}\,,\quad
  Q_\alpha =\hat{y}_\alpha\,,\quad
  S_\alpha =\hat{y}_\alpha k\,,\quad
  J=k+\nu \,.
$$
This observation guarantees that the system of equations
under consideration possesses N=2 global supersymmetry.
It is this $N=2$ supersymmetry which leads to a
doubled number of boson and fermion fields in the model.

\section{Bosonic Case and U(o(2,1))}

In the purely bosonic case one can proceed
in terms of bosonic operators,
avoiding the doubling of fields caused by supersymmetry.
To this end, let us use the orthogonal realization of the AdS algebra
$o(2,2)\sim o(2,1)\oplus o(2,1)$.
Let $T_a$ be the generators of $o(2,1)$,
\be
\label{comr}
   [T_a,T_b]=\epsilon_{ab}{}^c T_c \,,
\ee
where $\epsilon_{abc}$ is a totally antisymmetric 3d tensor,
$\epsilon_{012}=1$, and Latin indices are raised and lowered by
the Killing metrics of $o(2,1)$,
$$
   A^a=\eta^{ab}A_b\,,\qquad \eta=diag(1,-1,-1) \,.
$$

Let the background gravitational field have a form
\be
\label{W T}
   W_\mu=\omega_\mu{}^a T_a +\tilde\lambda\psi h_\mu{}^a T_a\,,
\ee
where $\psi$ is a central involutive element,
\be
   \psi^2=1,\qquad [\psi, T_a]=0\,,
\ee
and let $W$ obey the zero-curvature conditions (\ref{va}).
Note, that the inverse dreibein $h^\mu{}_a$
is normalized so that
\be
   h_\mu{}^a h^\mu{}^b=\eta^{ab} \,.
\ee
Let $T_a$ be restricted by the following additional condition on the
quadratic Casimir operator
\be
\label{tr}
   C_2\equiv T_a T^a=\frac18\left(\frac32-\frac{M^2}{\tilde\lambda^2}\right)\,.
\ee

We introduce the dynamical 0-form
$C$ as a function of $T_a$ and $\psi$
\be
\label{CT}
   C=\sum_{n=0}^\infty\sum_{A=0,1}\frac1{n!}\psi^A C_A{}^{a_1\ldots a_n}(x)
   T_{a_1}\ldots T_{a_n}\, ,
\ee
where $C_A{}^{a_1\ldots a_n}$ are totally symmetric traceless tensors.
Equivalently one can say that $C$ takes values in
the algebra $A_M \oplus A_M$ where
$A_M = U(o(2,1))/I_{(C_2-\frac18(\frac32-\frac{M^2}{\tilde\lambda^2}))}$.
Here $U(o(2,1))$ is the enveloping algebra for
the relations (\ref{comr}) and
$I_{(C_2-\frac18(\frac32-\frac{M^2}{\tilde\lambda^2}))}$ is the ideal
spanned by all elements of the form
$$
   \left[C_2-\frac18\left(\frac32-\frac{M^2}{\tilde\lambda^2}\right)\right]\,x
     \,,\qquad \forall x\in U(o(2,1) \,.
$$

We can then write down the equation analogous to~(\ref{D hatC})
in the form
\be
\label{DC T}
   D_{\mu}C=\tilde\lambda\psi h_{\mu}{}^a\{T_a,C\}\, ,
\ee
where
\be
   D_{\mu}C=\partial_{\mu}C-\omega_{\mu}{}^a[T_a,C]\, .
\ee
Acting on the both sides of eq.~(\ref{DC T}) by
the full covariant derivative $D^{\mu}$,
defined through the metric postulate
$D_{\mu}(h^a_{\nu}T_a)=0$
under the condition that the
Christoffel connection is symmetric, one can derive
\be
   \Box C_n=\frac12\tilde\lambda^2\left[2n(n+1)+\frac32-
    \frac{M^2}{\tilde\lambda^2} \right]C_n \,,  
\ee
where $C_n$ denotes a $n$-th power
monomial in (\ref{CT}). We see that this result coincides with
(\ref{L M}) at $N=2n$ and
\be
\label{ll}
     \lambda^2=\frac12\tilde\lambda^2 \,.
\ee

Also one can check that the zero-curvature conditions
for the gauge fields (\ref{W}) and (\ref{W T})
are equivalent to each other provided that (\ref{ll}) is true.
The explicit relationships are
$$
 \omega_\mu{}^{\alpha\beta}=-\frac12\omega_\mu{}^a\sigma_a^{\alpha\beta}\,,\quad
 h_\mu{}^{\alpha\beta}=-\frac1{\sqrt2}h_\mu{}^a\sigma_a^{\alpha\beta}\,,\quad
 T_a=-\frac1{16i}\sigma_a^{\alpha\beta}\{\hat{y}_\alpha,\hat{y}_\beta\}\,,
$$
where $\sigma_a^{\alpha\beta}=(I,\sigma_1,\sigma_3)$,
$\sigma_1\,,\sigma_3$ are symmetric Pauli matrices.

One can also check that, as expected, eq.~(\ref{DC T}) possesses
the same degenerate points in M as eq.~(\ref{DC mod}) does
according to~(\ref{a1}).

\section{Degenerate Points}

In this section we discuss briefly the specificities of
the equation~(\ref{D hatC}) at singular points in $\nu$.
Let us substitute the expansion~(\ref{hatC}) into~(\ref{DC mod})
with the coefficients defined by~(\ref{a})-(\ref{e})
and project (\ref{DC mod}) to the subspace of bosons $C_{+}$
by setting $k=1$ and $n$ to be even. Then we get in the component form
\be
\label{chain}
   DC_{\alpha(n)}=\frac i2\left[\left(1-\frac{\nu(\nu-2)}{(n+1)(n+3)}
   \right) h^{\beta\gamma}C_{\beta\gamma\alpha(n)}-
   \lambda^2n(n-1)h_{\alpha\alpha}C_{\alpha(n-2)}\right] \,.
\ee
In the general case (i.e., $\nu\ne 2l+1$, $l$-integer)
this chain of equations starts from the scalar component
and is equivalent to the dynamical equation~(\ref{M K-G})
with $M^2=\lambda^2\frac{\nu(\nu-2)}2$ supplemented either by
relations expressing highest multispinors via highest derivatives
of $C$ or identities which express the fact that higher derivatives
are symmetric.

At $\nu=2l+1$ the first term on the r.h.s.
of~(\ref{chain}) vanishes for $n=2(\pm l-1)$.
Since $n$ is non-negative let us choose for definiteness
a solution with $n=2(l-1)$, $l>0$.
One observes that the rank-$2l$ component is not any longer
expressed by~(\ref{chain}) via derivatives of the scalar $C$,
thus becoming an independent dynamical variable.
Instead, the equation (\ref{chain}) tells us that
(appropriately AdS covariantized) $l$-th derivative of the scalar field
$C$ vanishes. As a result, at degenerate points the system of equations
(\ref{chain}) acquires a non-decomposable triangle-type form with a
finite subsystem of equations for the set of
multispinors $C_{\alpha (2n)},$ $n<l$
and an infinite system of equations
for the dynamical field $C_{\alpha (2l)}$ and higher multispinors,
which contains (derivatives of)
the original field $C$ as a sort of sources on the right hand side.

The subsystem for lower multispinors describes a system analogous
to that of topological fields (\ref{aux}) which can contain at most a finite
number of degrees of freedom. In fact this system should be
dynamically trivial by the unitarity requirements (there are no finite-dimensional
unitary representations of the space-time symmetry groups)
\footnote{The only exception is when the degeneracy
takes place on the lowest level
and the representation turns out to be trivial
(constant).}.
Physically,
this is equivalent to imposing appropriate boundary conditions at infinity
which must kill these degrees of freedom because, having only a finite number
of non-vanishing derivatives, these fields have a polynomial growth
at the space-time infinity (except for a case of a constant field $C$).

Thus one can factor out the decoupling lowest components arriving
at the system of equations which starts from the field $C_{\alpha (2l)}$.
These systems are dynamically non-trivial and correspond to
certain gauge systems. For example, one can show that the first degenerate
point $\nu=3$ just corresponds to 3d electrodynamics.
To see this one can introduce a two-form
\be
   F=h^{\alpha}{}_{\gamma}\wedge h^{\gamma\beta}
   C_{\alpha\beta}\,
\ee
and verify that the infinite part of the system (\ref{chain}) with
$n\ge2$ (i.e. with the scalar field factored out) is equivalent to the
Maxwell equations
\be
   dF=0\,,\qquad d\,{}^* F=0\,
\ee
supplemented with an infinite chain of higher Bianchi identities
(here ${}^* F$ denotes a form dual to $F$).
Note that, for our normalization of a mass,
electrodynamics turns out to be
massive with the mass $M^2=\frac32\lambda^2$ which
vanishes in the flat limit $\lambda\to 0 $.
A more detailed analysis of this formulation of
electrodynamics and its counterparts corresponding to
higher degenerate points will be given in \cite{Fut}.

Now let us note that there exists an alternative formulation of
the dynamics of matter fields which is equivalent to the original one
of \cite{Unf} for all $\nu$ and is based on the co-twisted representation
$\tilde C$ . Namely, let us introduce a non-degenerate invariant
form
\be
  \langle C,\tilde C \rangle = \int d^4x \sum_{n=0}^\infty
  \frac 1{(2n)!}C_{\alpha(2n)}\tilde C^{\alpha(2n)}\,
\ee
confining ourselves for simplicity to the purely bosonic case
in the sector $C_{+}$.

The covariant differential corresponding to the
twisted representation $C$ of $o(2,2)$ has the form
\be
   {\cal D}C= dC-[\omega,C]-\lambda\{h,C\}\, ,
\ee
so that eq.~(\ref{D hatC}) acquires a form \quad
${\cal D}C=0$.
The covariant derivative in the co-twisted representation can be obtained
from the invariance condition
\be
  \langle C,{\cal D}\tilde C \rangle =-\langle {\cal D}C,\tilde C \rangle
   \,.
\ee
It has the following explicit form
\begin{eqnarray}
   \lefteqn{{\cal D}\tilde C^{\alpha(n)}=d\tilde C^{\alpha(n)}-
     n\omega^{\alpha}{}_{\beta}\tilde C^{\beta\alpha(n-1)} }\nonumber\\
     & & {}-\frac i2\left[h_{\beta\gamma}\tilde C^{\beta\gamma\alpha(n)}-
     \lambda^2n(n-1)\left(1-\frac{\nu(\nu-2)}{(n-1)(n+1)}\right)
     h^{\alpha\alpha}\tilde C^{\alpha(n-2)}\right] \,.
\end{eqnarray}
As a result the equation for $\tilde C$ analogous
to~(\ref{chain}) reads
\be
\label{co-chain}
   D\tilde C^{\alpha(n)}=\frac i2\left[
   h_{\beta\gamma}\tilde C^{\beta\gamma\alpha(n)}-
   \lambda^2n(n-1)\left(1-\frac{\nu(\nu-2)}{(n-1)(n+1)}\right)
   h^{\alpha\alpha}\tilde C^{\alpha(n-2)}\right] \,.
\ee

We see that now the term containing a higher multispinor
appears with a unite coefficient while the coefficients in front of the
lower multispinor sometimes vanish.
The equations (\ref{co-chain}) identically
coincide with the equations derived in \cite{Unf}
which are reproduced in the section 2 of this paper.
Let us note that the twisted and co-twisted representations are equivalent for
all $\nu \neq 2l+1$ because the algebra of deformed oscillators
possesses an invariant quadratic form which is non-degenerate
for all $\nu \neq 2l+1$ \cite{Quant}.
For $\nu = 2l+1$ this is not the case
any longer since the invariant quadratic form degenerates and therefore
twisted and co-twisted representations turn out to be formally inequivalent.

Two questions are now in order. First, what is a physical difference
between the equations corresponding to twisted and co-twisted representations
at the degenerate points, and second which of these two representations can be used
in an interacting theory. These issues will be considered in more detail
in \cite{Fut}. Here we just mention that at the free field level the two
formulations are still physically equivalent and in fact turn out to be dual to each other.
For example for the case of electrodynamics the scalar field component $C$
in the co-twisted representation can be interpreted as
a magnetic potential such that ${}^*F = dC$. A non-trivial question
then is whether such a formulation can be extended
to any consistent local interacting theory. Naively one can expect that the formulation
in terms of the twisted representation has better chances to be extended beyond the
linear problem. It will be shown in \cite{Fut} that this is indeed the case.

\section{Conclusion}

In this paper we suggested a simple algebraic
method of  formulating free field equations
for massive spin-0 and spin 1/2 matter fields
in 2+1 dimensional AdS space in the form of covariant
constantness conditions for certain infinite-dimensional
representations of the space-time symmetry group.
An important advantage of this formulation is that it
allows one to describe in a simple way a structure of the global
higher-spin symmetries. These symmetries are described by the
parameters which take values in the infinite-dimensional
algebra of functions of all generating elements $y_\alpha$, $k$
and $\psi$, i.e.
$\varepsilon=\varepsilon(y_\alpha,k,\psi |x)$. The full
transformation law has a form
\be
\label{trans}
   \delta C = \varepsilon C - C\tilde\varepsilon\,,
\ee
where
\be
   \tilde\varepsilon(y_\alpha,k,\psi |x)=\varepsilon(y_\alpha,k,-\psi |x)
\ee
and the dependence of
$\varepsilon $ on $x$ is fixed by the equation
\be
   d \varepsilon=W_{gr} \varepsilon  - \varepsilon W_{gr} \,,
\ee
which is integrable as a consequence of the zero-curvature
conditions (\ref{va}) and
therefore admits a unique solution in terms of an arbitrary
function $\varepsilon_0 (y_\alpha,k,\psi)=\varepsilon
(y_\alpha,k,\psi|x_0)$ for an arbitrary point of space-time
$x_0$.  It is obvious that the equations (\ref{D hatC})
are indeed invariant with respect to the transformations
(\ref{trans}).

Explicit knowledge of the structure of the global higher-spin symmetry
is one of the main results obtained in this paper. In \cite{Fut}
it will serve as a starting point for the analysis of higher-spin
interactions of matter fields in 2+1 dimension. An interesting feature of higher-spin
symmetries demonstrated in this paper
is that their form depends on a particular dynamical system under
consideration. Indeed, the higher-spin algebras with
different $M^2 (\nu )$ are pairwise non-isomorphic.
This is obvious from the identification of the higher-spin symmetries with
certain factor-algebras of
the enveloping algebras of space-time symmetry algebras
along the lines of the Section 4. Ordinary space-time symmetries on their turn
can be identified with (maximal)
finite-dimensional subalgebras of the higher-spin algebras which do not
depend on the dynamical parameters like $\nu$ (cf (\ref{y mod})).

The infinite-dimensional algebras isomorphic to
those considered in the section 4 have been originally introduced
in \cite{BBS,H}
as candidates for 3d  bosonic higher-spin algebras, while the
superalgebras of deformed oscillators described in the section 3
were suggested in \cite{Quant} as candidates for 3d higher-spin
superalgebras. Using all these algebras
and the definition of supertrace given in \cite{Quant}
it was possible to write
a Chern-Simons action for the 3d higher-spin gauge fields which
are all dynamically trivial in the absence of matter fields
(in a topologically trivial situation).  Originally this was done
by Blencowe \cite{bl} for the case of the Heisenberg algebra (i.e. $\nu =0$).
It was not clear, however, what is a physical meaning of the
ambiguity in the continuous parameter like $\nu$ parametrizing
pairwise non-isomorphic 3d higher-spin algebras. In this paper
we have shown that different symmetries are realized on different
matter multiplets, thus concluding that higher-spin symmetries
turn out to be dependent on a particular physical model
under consideration.

\section*{Acknowledgements}

The research described in this article
 was supported in part by the European Community
Commission under the contract INTAS, Grant No.94-2317 and by the
Russian Foundation for Basic Research, Grant No.96-01-01144.


\begin{thebibliography}{99}

\bibitem{Unf} M.A.~Vasiliev, Class.~Quant.~Grav. {\bf 11} (1994) 649.
\bibitem{Eq} M.A.~Vasiliev, Mod.~Phys.~Lett. {\bf A7} (1992) 3689.
\bibitem{Quant} M.A.~Vasiliev,
  JETP Lett. {\bf 50} (1989) N8, 374;
Int. J. Mod. Phys. {\bf A6} (1991) 1115.
\bibitem{Fut} S.F.~Prokushkin and M.A.~Vasiliev, in preparation.
\bibitem{Wig} E.P.~Wigner, Phys. Rev. {\bf 77} (1950) 711.
\bibitem{BWV}
E.~Bergshoeff, B.~de~Wit and M.~A.~Vasiliev,
               Nucl.~Phys. {\bf B366} (1991) 315.
\bibitem{BBS}
E.~Bergshoeff, M.P.~Blencowe and K.S.~Stelle, Commun.~Math.~Phys. {\bf 128}
(1990) 213.
\bibitem{H}
M.~Bordemann, J.~Hoppe and P.~Schaller, Phys.~Lett. {\bf 232} (1989) 199.
\bibitem{bl}
M.P.~Blencowe, Class. Quantum Grav. {\bf 6} (1989) 443.
\end{thebibliography}
\end{document}